\documentclass[submission,copyright,creativecommons]{eptcs}
\providecommand{\event}{UITP 2014} 

\usepackage{graphicx}

\usepackage{amsmath, amsthm, color}
\definecolor{light-gray}{gray}{0.75}

\usepackage{wrapfig}
\usepackage{tikzfig}
\tikzstyle{dotpic}=[scale=0.6]
\tikzstyle{diredges}=[every to/.style={diredge}]
\tikzstyle{dot graph}=[shorten <=-0.1mm,shorten >=-0.1mm,scale=0.6]
\tikzstyle{digraph}=[-latex]
\tikzstyle{plot point}=[circle,fill=black,minimum width=2mm,inner sep=0]
\tikzstyle{string graph}=[]
\tikzstyle{every picture}=[string graph,baseline=-0.25em,scale=0.6]
\tikzstyle{sg diredge}=[-stealth]
\tikzstyle{rewrite edge}=[-open triangle 45]
\tikzstyle{sg bold diredge}=[-stealth,thick,shorten >=-1pt]
\tikzstyle{sg vertex}=[circle,minimum width=2.2mm,fill=white,draw=black,inner sep=0mm]
\tikzstyle{labelled sg vertex}=[circle,minimum width=7mm,fill=white,draw=black,inner sep=0mm]
\tikzstyle{small sg vertex}=[circle,minimum width=4mm,fill=white,draw=black,inner sep=0mm, font=\footnotesize]
\tikzstyle{sg grey vertex}=[sg vertex,fill=gray!30!white]
\tikzstyle{sg white vertex}=[sg vertex,fill=white]
\tikzstyle{sg black vertex}=[sg vertex,fill=black]
\tikzstyle{sg bold vertex}=[circle,minimum width=2.2mm,fill=white,draw=black,very thick,inner sep=0mm]
\tikzstyle{sg wire vertex}=[circle,minimum width=1mm,fill=black,inner sep=0mm]
\tikzstyle{cloud vertex}=[fill=white, draw=black, inner sep=2 mm, shape=cloud, aspect=1.5]
\tikzstyle{tick vertex}=[rectangle,fill=black,minimum height=1mm,minimum width=2.5mm,inner sep=0mm]

\tikzstyle{braceedge}=[decorate,decoration={brace,amplitude=2mm,raise=-1mm}]
\tikzstyle{small braceedge}=[decorate,decoration={brace,amplitude=1mm,raise=-1mm}]
\tikzstyle{left hook arrow}=[left hook-latex]
\tikzstyle{right hook arrow}=[right hook-latex]

\tikzstyle{bbox edge}=[draw=blue]
\tikzstyle{bbox include}=[->,draw=blue]
\tikzstyle{bbox corner}=[inner sep=0pt,rectangle,fill=blue,draw=blue,minimum width=1.5mm,minimum height=1.5mm]

\tikzstyle{west wire label}=[font=\footnotesize\it,anchor=west,inner sep=1pt,xshift=-3pt]
\tikzstyle{east wire label}=[font=\footnotesize\it,anchor=east,inner sep=1pt,xshift=3pt]

\tikzstyle{dot}=[inner sep=0.7mm,minimum width=0pt,minimum height=0pt,fill=black,draw=black,shape=circle]
\tikzstyle{white dot}=[dot,fill=white]
\tikzstyle{alt white dot}=[white dot,label={[xshift=2.9mm,yshift=-0.1mm]left:$\cdot$}]
\tikzstyle{gray dot}=[dot,fill=gray!50]
\tikzstyle{box vertex}=[draw=black,rectangle]
\tikzstyle{whitebg}=[fill=white,inner sep=2pt]
\tikzstyle{graph state vertex}=[sg vertex,fill=black]

\tikzstyle{square box}=[rectangle,fill=white,draw=black,minimum height=6mm,minimum width=6mm]
\tikzstyle{small square box}=[rectangle,fill=white,draw=black,minimum height=0mm,minimum width=8mm,inner sep=3pt]
\tikzstyle{square gray box}=[rectangle,fill=gray!30,draw=black,minimum height=6mm,minimum width=6mm]
\tikzstyle{point}=[regular polygon,regular polygon sides=3,draw=black,scale=0.75,inner sep=-0.5pt,minimum width=7mm,fill=white]
\tikzstyle{copoint}=[point,regular polygon rotate=180,fill=white]
\tikzstyle{gray point}=[point,fill=gray!40!white]
\tikzstyle{gray copoint}=[copoint,fill=gray!40!white]

\tikzstyle{open graph}=[baseline=-0.25em]
\tikzstyle{greybg}=[background rectangle/.style={fill=black!5,draw=black!30,rounded corners=1ex}, show background rectangle]

\tikzstyle{edge point}=[circle,minimum width=1mm,fill=black,inner sep=0mm]
\tikzstyle{vertex point}=[circle,minimum width=2.2mm,fill=white,draw=black,inner sep=0mm]
\tikzstyle{gray vertex point}=[circle,minimum width=2.2mm,fill=gray!30!white,draw=black,inner sep=0mm]
\tikzstyle{edge label}=[inner sep=2pt, font=\small]
\tikzstyle{on edge label}=[fill=white, font=\footnotesize, inner sep=1 pt]

\newcommand{\edgearrow}{{\arrow[black]{>}}}
\newcommand{\edgetick}{{\arrow[black,scale=0.7,very thick]{|}}}
\tikzstyle{diredge}=[->]
\tikzstyle{gray edge}=[gray!50!white]
\tikzstyle{medium diredge}=[->]
\tikzstyle{short diredge}=[->]
\tikzstyle{halfedge}=[-)]
\tikzstyle{other halfedge}=[(-]
\tikzstyle{freeedge}=[(-)]
\tikzstyle{white edge}=[line width=5pt,white]
\tikzstyle{tick}=[postaction=decorate,decoration={markings, mark=at position 0.5 with \edgetick}]
\tikzstyle{small map edge}=[|-latex, gray!60!blue, shorten <=0.9mm, shorten >=0.5mm]
\tikzstyle{thick dashed edge}=[very thick,dashed,gray!40]
\tikzstyle{map edge}=[|-latex,very thick, gray!40, shorten <=1mm, shorten >=0.5mm]
\tikzstyle{tickedge}=[postaction=decorate,
  decoration={markings, mark=at position 0.5 with \edgetick}]
\tikzstyle{dirtickedge}=[postaction=decorate,
  decoration={markings, mark=at position 0.5 with \edgetick},
  decoration={markings, mark=at position 0.85 with \edgearrow}]
\tikzstyle{dirdoubletickedge}=[postaction=decorate,
  decoration={markings, mark=at position 0.4 with \edgetick},
  decoration={markings, mark=at position 0.6 with \edgetick},
  decoration={markings, mark=at position 0.85 with \edgearrow}]

\tikzstyle{arrs}=[-latex,font=\small,auto]
\tikzstyle{arrow plain}=[arrs]
\tikzstyle{arrow dashed}=[dashed,arrs]
\tikzstyle{arrow bold}=[very thick,arrs]
\tikzstyle{arrow hide}=[draw=white!0,-]
\tikzstyle{arrow reverse}=[latex-]
\tikzstyle{cdnode}=[]

\tikzstyle{cnot}=[fill=white,shape=circle,inner sep=-1.4pt]
\tikzstyle{bang box}=[draw=black,dashed,minimum height=12mm,minimum width=12mm,fill=gray!20]
\tikzstyle{wire label}=[font=\footnotesize, auto]

\newcommand{\cmdrewritesto}{\tikz[baseline=-0.25em] { \draw [-open triangle 45, line width=0.2pt] (0,0) -- (0.5,0); }\,}

\DeclareMathOperator{\rewritesto}{\cmdrewritesto}

\tikzstyle{cdiag}=[matrix of math nodes, row sep=2em, column sep=2em, text height=1.5ex, text depth=0.25ex,inner sep=0.5em]
\tikzstyle{arrow above}=[transform canvas={yshift=0.5ex}]
\tikzstyle{arrow below}=[transform canvas={yshift=-0.5ex}]

\title{Tinker, tailor, solver, proof\thanks{This work has been supported by EPSRC grants: EP/H023852, EP/H024204 and EP/J001058, the John Templeton Foundation, and the Office of Navel Research.}}
\author{Gudmund Grov
\institute{Heriot-Watt University, Edinburgh, UK} 
\email{G.Grov@hw.ac.uk}
\and
Aleks Kissinger
\institute{University of Oxford, UK} 
\email{alek@cs.ox.ac.uk}
\and 
Yuhui Lin
\institute{Heriot-Watt University, Edinburgh, UK} 
\email{Y.Lin@hw.ac.uk}
}
\def\titlerunning{Tinker, tailor, solver, proof}
\def\authorrunning{G. Grov, A. Kissinger and Y. Lin}
\begin{document}
\maketitle

\begin{abstract}
We introduce Tinker, a tool for designing and evaluating proof strategies based on \textit{proof-strategy graphs}, a formalism introduced by the authors in~\cite{LPAR13}. We represent proof strategies as open-graphs, which are directed graphs with additional input/output edges. Tactics appear as nodes in a graph, and can be `piped' together by adding edges between them. Goals are added to the input edges of such a graph, and flow through the graph as the strategy is evaluated. Properties of the edges ensure that only the right `type' of goals are accepted. In this paper, we detail the Tinker tool and show how it can be integrated with two different theorem provers: Isabelle and ProofPower.
\end{abstract}

\section{Proof strategy graphs -- a framework for tinkering}

Traditionally, the way to build up a proof from atomic tactics is to chain them together (possibly with some repetition, alternation, etc.), letting later tactics act on sub-goals produced by earlier ones. However, without any extra `plumbing' to control the flow of goals through a proof, one tends to depend on semantically irrelevant data, such as the \textit{order} in which tactics produce goals to dictate where those goals might end up, which leads to brittleness in proof strategies. For example, consider a case where we expect three sub-goals from tactic $t_1$, where the first two are sent to $t_2$ and the last to $t_3$. A small improvement of $t_1$ may result in only two sub-goals. This ``improvement'' causes $t_2$ to be applied to the second goal when it should have been $t_3$. The tactic $t_2$ may then fail or create unexpected new sub-goals that cause some later tactic to fail.

\begin{figure}
  \centering
  \scalebox{1.0}{%
\beginpgfgraphicnamed{rewrite_ex}
\begin{tikzpicture}[string graph]
	\begin{pgfonlayer}{nodelayer}
		\node [style=square box] (0) at (-9.25, -0.25) {ripple};
		\node [style=none] (1) at (-8.75, -1.25) {};
		\node [style=small sg vertex] (2) at (-10.5, 3) {a};
		\node [style=none] (3) at (-10, 1) {};
		\node [style=none] (4) at (-10.5, 3.5) {};
		\node [style=none] (5) at (-9.75, 0.25) {};
		\node [style=square box] (6) at (-10.5, 2) {induct};
		\node [style=none] (7) at (-9.75, -2.75) {};
		\node [style=none] (8) at (-11, -0.5) {};
		\node [style=none] (9) at (-11, -2.75) {};
		\node [style=none] (10) at (-10.5, 2.5) {};
		\node [style=none] (11) at (-11, 0.25) {};
		\node [style=none] (12) at (-9.75, -0.75) {};
		\node [style=none] (13) at (-7.75, 0.75) {};
		\node [style=none] (14) at (-8.75, 0.75) {};
		\node [style=none] (15) at (-7.75, -1.25) {};
		\node [style=none] (16) at (-7, 0) {$\rewritesto$};
		\node [style=none] (17) at (-8.75, -0.75) {};
		\node [style=none] (18) at (-10, 1.5) {};
		\node [style=none] (19) at (-11, 1.5) {};
		\node [style=none] (20) at (-8.75, 0.25) {};
		\node [style=none] (21) at (-5.75, -2.75) {};
		\node [style=none] (22) at (-4.5, -2.75) {};
		\node [style=none] (23) at (-5.25, 3.5) {};
		\node [style=none] (24) at (-3.5, 0.75) {};
		\node [style=none] (25) at (-5.75, 0.25) {};
		\node [style=none] (26) at (-3.5, -0.75) {};
		\node [style=small sg vertex] (27) at (-5.75, 0.25) {b};
		\node [style=none] (28) at (-4.5, 0.25) {};
		\node [style=none] (29) at (-3.5, 0.25) {};
		\node [style=none] (30) at (-1.75, 0) {$\rewritesto$};
		\node [style=none] (31) at (-2.5, -1.25) {};
		\node [style=square box] (32) at (-4, -0.25) {ripple};
		\node [style=square box] (33) at (-5.25, 2) {induct};
		\node [style=none] (34) at (-3.5, -1.25) {};
		\node [style=none] (35) at (-4.75, 0.75) {};
		\node [style=none] (36) at (-5.25, 2.5) {};
		\node [style=none] (37) at (-4.75, 1.5) {};
		\node [style=small sg vertex] (38) at (-4.75, 1) {d};
		\node [style=none] (39) at (-5.75, 1.5) {};
		\node [style=none] (40) at (-4.5, -0.75) {};
		\node [style=none] (41) at (-5.75, -0.5) {};
		\node [style=none] (42) at (-2.5, 0.75) {};
		\node [style=small sg vertex] (43) at (-0.5, -1.25) {c};
		\node [style=none] (44) at (-0.5, -2.75) {};
		\node [style=none] (45) at (0.75, -2.75) {};
		\node [style=none] (46) at (0, 3.5) {};
		\node [style=none] (47) at (1.75, 0.75) {};
		\node [style=none] (48) at (-0.5, 1) {};
		\node [style=none] (49) at (1.75, -0.75) {};
		\node [style=small sg vertex] (50) at (0.75, -1.25) {d};
		\node [style=none] (51) at (0.75, 0.25) {};
		\node [style=none] (52) at (1.75, 0.25) {};
		\node [style=none] (53) at (3.5, 0) {$\rewritesto$};
		\node [style=none] (54) at (2.75, -1.25) {};
		\node [style=square box] (55) at (1.25, -0.25) {ripple};
		\node [style=square box] (56) at (0, 2) {induct};
		\node [style=none] (57) at (1.75, -1.25) {};
		\node [style=none] (58) at (0.5, 1) {};
		\node [style=none] (59) at (0, 2.5) {};
		\node [style=none] (60) at (0.5, 1.5) {};
		\node [style=small sg vertex] (61) at (1.75, -1.25) {e};
		\node [style=none] (62) at (-0.5, 1.5) {};
		\node [style=none] (63) at (0.75, -0.75) {};
		\node [style=none] (64) at (-0.5, 0.25) {};
		\node [style=none] (65) at (2.75, 0.75) {};
		\node [style=small sg vertex] (66) at (4.75, -1.25) {c};
		\node [style=none] (67) at (4.75, -2.75) {};
		\node [style=none] (68) at (6, -2.75) {};
		\node [style=none] (69) at (5.25, 3.5) {};
		\node [style=none] (70) at (7, 0.75) {};
		\node [style=none] (71) at (4.75, 1) {};
		\node [style=none] (72) at (7, -0.75) {};
		\node [style=small sg vertex] (73) at (6, -2) {d};
		\node [style=none] (74) at (6, 0.25) {};
		\node [style=none] (75) at (7, 0.25) {};
		\node [style=none] (76) at (8.75, 0) {$\rewritesto$};
		\node [style=none] (77) at (8, -1.25) {};
		\node [style=square box] (78) at (6.5, -0.25) {ripple};
		\node [style=square box] (79) at (5.25, 2) {induct};
		\node [style=none] (80) at (7, -1.25) {};
		\node [style=none] (81) at (5.75, 1) {};
		\node [style=none] (82) at (5.25, 2.5) {};
		\node [style=none] (83) at (5.75, 1.5) {};
		\node [style=small sg vertex] (84) at (7, 0.75) {e};
		\node [style=none] (85) at (4.75, 1.5) {};
		\node [style=none] (86) at (6, -0.75) {};
		\node [style=none] (87) at (4.75, 0.25) {};
		\node [style=none] (88) at (8, 0.75) {};
		\node [style=none] (89) at (13.25, 0.75) {};
		\node [style=none] (90) at (12.25, 0.25) {};
		\node [style=none] (91) at (11, 1.5) {};
		\node [style=none] (92) at (11.25, 0.25) {};
		\node [style=small sg vertex] (93) at (10, -1.25) {c};
		\node [style=small sg vertex] (94) at (11.25, -1.25) {f};
		\node [style=none] (95) at (10, 0.25) {};
		\node [style=none] (96) at (10.5, 2.5) {};
		\node [style=square box] (97) at (10.5, 2) {induct};
		\node [style=none] (98) at (10, 1) {};
		\node [style=none] (99) at (10, 1.5) {};
		\node [style=small sg vertex] (100) at (11.25, -2) {d};
		\node [style=none] (101) at (11, 1) {};
		\node [style=square box] (102) at (11.75, -0.25) {ripple};
		\node [style=none] (103) at (11.25, -0.75) {};
		\node [style=none] (104) at (12.25, -0.75) {};
		\node [style=none] (105) at (10, -2.75) {};
		\node [style=none] (106) at (11.25, -2.75) {};
		\node [style=none] (107) at (12.25, -1.25) {};
		\node [style=none] (108) at (10.5, 3.5) {};
		\node [style=none] (109) at (12.25, 0.75) {};
		\node [style=none] (110) at (13.25, -1.25) {};
		\node [style=small sg vertex] (111) at (10, -2) {b};
		\node [style=small sg vertex] (112) at (4.75, -2) {b};
		\node [style=small sg vertex] (113) at (-0.5, -2) {b};
		\node [style=small sg vertex] (114) at (-5.75, 1) {c};
	\end{pgfonlayer}
	\begin{pgfonlayer}{edgelayer}
		\draw [style=diredge] (4.center) to (10.center);
		\draw [style=diredge, in=90, out=-90, looseness=1.25] (3.center) to (5.center);
		\draw [style=diredge, in=90, out=-90] (12.center) to (7.center);
		\draw [in=90, out=-90, looseness=1.25] (11.center) to (8.center);
		\draw [style=diredge] (8.center) to (9.center);
		\draw [in=-90, out=-90, looseness=2.00] (1.center) to (15.center);
		\draw (15.center) to (13.center);
		\draw [in=90, out=90, looseness=2.00] (13.center) to (14.center);
		\draw (17.center) to (1.center);
		\draw (18.center) to (3.center);
		\draw (19.center) to (11.center);
		\draw [style=diredge] (14.center) to (20.center);
		\draw [style=diredge] (23.center) to (36.center);
		\draw [style=diredge, in=90, out=-90, looseness=1.25] (35.center) to (28.center);
		\draw [style=diredge, in=90, out=-90] (40.center) to (22.center);
		\draw [in=90, out=-90, looseness=1.25] (25.center) to (41.center);
		\draw [style=diredge] (41.center) to (21.center);
		\draw [in=-90, out=-90, looseness=2.00] (34.center) to (31.center);
		\draw (31.center) to (42.center);
		\draw [in=90, out=90, looseness=2.00] (42.center) to (24.center);
		\draw (26.center) to (34.center);
		\draw (37.center) to (35.center);
		\draw (39.center) to (25.center);
		\draw [style=diredge] (24.center) to (29.center);
		\draw [style=diredge] (46.center) to (59.center);
		\draw [style=diredge, in=90, out=-90, looseness=1.25] (58.center) to (51.center);
		\draw [style=diredge, in=90, out=-90] (63.center) to (45.center);
		\draw [in=90, out=-90, looseness=1.25] (48.center) to (64.center);
		\draw [style=diredge] (64.center) to (44.center);
		\draw [in=-90, out=-90, looseness=2.00] (57.center) to (54.center);
		\draw (54.center) to (65.center);
		\draw [in=90, out=90, looseness=2.00] (65.center) to (47.center);
		\draw (49.center) to (57.center);
		\draw (60.center) to (58.center);
		\draw (62.center) to (48.center);
		\draw [style=diredge] (47.center) to (52.center);
		\draw [style=diredge] (69.center) to (82.center);
		\draw [style=diredge, in=90, out=-90, looseness=1.25] (81.center) to (74.center);
		\draw [style=diredge, in=90, out=-90] (86.center) to (68.center);
		\draw [in=90, out=-90, looseness=1.25] (71.center) to (87.center);
		\draw [style=diredge] (87.center) to (67.center);
		\draw [in=-90, out=-90, looseness=2.00] (80.center) to (77.center);
		\draw (77.center) to (88.center);
		\draw [in=90, out=90, looseness=2.00] (88.center) to (70.center);
		\draw (72.center) to (80.center);
		\draw (83.center) to (81.center);
		\draw (85.center) to (71.center);
		\draw [style=diredge] (70.center) to (75.center);
		\draw [style=diredge] (108.center) to (96.center);
		\draw [style=diredge, in=90, out=-90, looseness=1.25] (101.center) to (92.center);
		\draw [style=diredge, in=90, out=-90] (103.center) to (106.center);
		\draw [in=90, out=-90, looseness=1.25] (98.center) to (95.center);
		\draw [style=diredge] (95.center) to (105.center);
		\draw [in=-90, out=-90, looseness=2.00] (107.center) to (110.center);
		\draw (110.center) to (89.center);
		\draw [in=90, out=90, looseness=2.00] (89.center) to (109.center);
		\draw (104.center) to (107.center);
		\draw (91.center) to (101.center);
		\draw (99.center) to (98.center);
		\draw [style=diredge] (109.center) to (90.center);
	\end{pgfonlayer}
\end{tikzpicture}}
\endpgfgraphicnamed}
  \caption{Some goal-nodes (depicted as circles) making their way through a PSGraph}\label{fig:ripple-ex}
\end{figure}
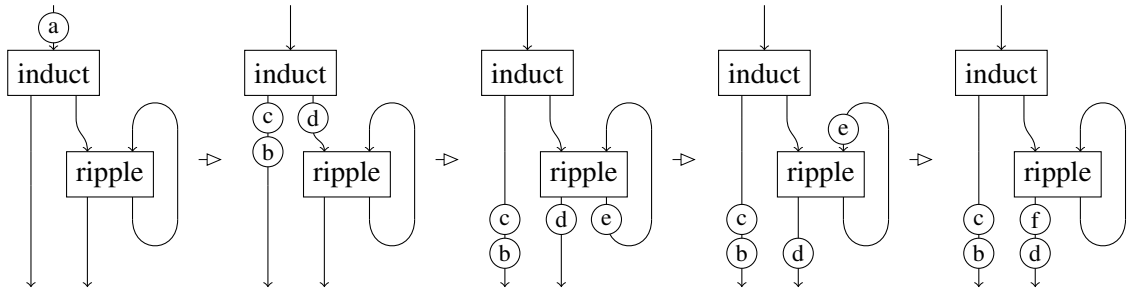

There are a handful of ways one might go about adding this extra plumbing to a proof strategy. The technique provided by the Tinker tool\footnote{
Available at \url{https://github.com/ggrov/psgraph/tree/uitp14}.}, detailed in this paper, makes use of \textit{proof-strategy graphs} (PSGraphs), as defined in~\cite{LPAR13}, which take the notion of goal-plumbing quite literally. Tactics appear as nodes in an open-graph, which is essentially a directed graph with the additional property that we allow dangling edges, i.e. edges without source and/or target nodes. Edges without source or target nodes serve as inputs and outputs to the graph as a whole. One evaluates a PSGraph by placing one or more goal-nodes, each containing a single goal, on input edges of the graph, then applying tactics to consume goals on the in-edges of a tactic-node and producing sub-goals on the out-edges. As a result, the goals appear to flow through the graph, hitting tactics along the way, until they are either consumed (i.e. closed), or reach the output edges of the graph, in which case they become output goals for the overall evaluation. In Figure ~\ref{fig:ripple-ex}, we show some goals making their way through a PSGraph with two nodes, labelled by the tactics they represent (in this case `induct' and `ripple').

Note how a single tactic node can have many out-edges. In principle, applying a tactic could send sub-goals to any of the output edges. However, there will typically only be one particular output that is appropriate for a any particular goal. When plumbing a house, pipes comes in all sizes and shapes, and you can only connect the same types of pipes together. The same is true for tactics: they only work for certain goals (although for some tactics this range of goals is rather wide). For example, an `assumption' tactic expects a hypothesis to be unifiable with the goal, and `$\forall$-intro' expects the goal to start with a $\forall$ quantifier. 

\begin{wrapfigure}[6]{r}{0.2\textwidth}
  \centering
\beginpgfgraphicnamed{induct}
\begin{tikzpicture}
	\begin{pgfonlayer}{nodelayer}
		\node [style=west wire label] (0) at (0.25, 1) {inductable};
		\node [style=west wire label] (1) at (0.75, -1) {step};
		\node [style=none] (2) at (0, 1.25) {};
		\node [style=none] (3) at (0, 0.5) {};
		\node [style=none] (4) at (0.5, -1) {};
		\node [style=square box] (5) at (0, 0) {induct};
		\node [style=none] (6) at (0.5, -0.5) {};
		\node [style=none] (7) at (1.25, -1.75) {};
		\node [style=east wire label] (8) at (-0.75, -1) {base};
		\node [style=none] (9) at (-0.5, -0.5) {};
		\node [style=none] (10) at (-1.25, -1.75) {};
		\node [style=none] (11) at (-0.5, -1) {};
	\end{pgfonlayer}
	\begin{pgfonlayer}{edgelayer}
		\draw [style=diredge, in=90, out=-90, looseness=1.25] (4.center) to (7.center);
		\draw (6.center) to (4.center);
		\draw [style=diredge] (2.center) to (3.center);
		\draw [style=diredge, in=90, out=-90, looseness=1.25] (11.center) to (10.center);
		\draw (9.center) to (11.center);
	\end{pgfonlayer}
\end{tikzpicture}}
\endpgfgraphicnamed
\end{wrapfigure}
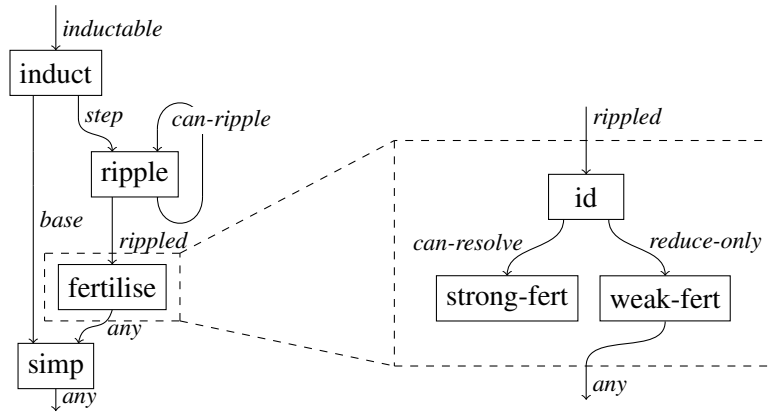
\begin{figure}[h]
  \centering
\beginpgfgraphicnamed{unfold_ex}
\begin{tikzpicture}[string graph]
	\begin{pgfonlayer}{nodelayer}
		\node [style=west wire label, fill=white, xshift=3pt, yshift=-1pt] (0) at (1.75, 2) {can-ripple};
		\node [style=none] (1) at (1.5, 1.25) {};
		\node [style=none] (2) at (-1.25, 1.25) {};
		\node [style=none] (3) at (1.5, 1.75) {};
		\node [style=square box] (4) at (-0.75, 3) {induct};
		\node [style=none] (5) at (1.5, 0.25) {};
		\node [style=none] (6) at (-1.25, -3) {};
		\node [style=none] (7) at (0.5, 1.25) {};
		\node [style=square box] (8) at (1, 0.75) {ripple};
		\node [style=none] (9) at (-0.25, 2.5) {};
		\node [style=none] (10) at (2.5, 0.25) {};
		\node [style=none] (11) at (0.5, -1.25) {};
		\node [style=none] (12) at (2.5, 1.75) {};
		\node [style=none] (13) at (-1.25, 0.5) {};
		\node [style=none] (14) at (0.5, 0.25) {};
		\node [style=none] (15) at (-1.25, 2.5) {};
		\node [style=none] (16) at (1.5, 0.25) {};
		\node [style=none] (17) at (-0.25, 2) {};
		\node [style=west wire label] (18) at (0, 2) {step};
		\node [style=west wire label] (19) at (0.75, -0.75) {rippled};
		\node [style=square box] (20) at (0.5, -1.75) {fertilise};
		\node [style=west wire label] (21) at (-0.5, -4.25) {any};
		\node [style=square box] (22) at (-0.75, -3.5) {simp};
		\node [style=none] (23) at (-0.75, -4.5) {};
		\node [style=none] (24) at (-0.75, -4) {};
		\node [style=none] (25) at (0.5, -2.25) {};
		\node [style=none] (26) at (-0.25, -3) {};
		\node [style=west wire label] (27) at (0.5, -2.75) {any};
		\node [style=west wire label] (28) at (-1, -0.25) {base};
		\node [style=west wire label] (29) at (-0.5, 4) {inductable};
		\node [style=none] (30) at (-0.75, 3.5) {};
		\node [style=none] (31) at (-0.75, 4.5) {};
		\node [style=none] (32) at (9.25, -1.5) {};
		\node [style=none] (33) at (11, 0.75) {};
		\node [style=square box] (34) at (12.75, -2) {weak-fert};
		\node [style=none] (35) at (11, -3.5) {};
		\node [style=east wire label] (36) at (9.5, -0.75) {can-resolve};
		\node [style=west wire label] (37) at (11.25, -4) {any};
		\node [style=none] (38) at (11, 2.25) {};
		\node [style=none] (39) at (11, -4.25) {};
		\node [style=none] (40) at (12.75, -1.5) {};
		\node [style=none] (41) at (11.5, -0.25) {};
		\node [style=square box, minimum width=1 cm] (42) at (11, 0.25) {id};
		\node [style=none] (43) at (10.5, -0.25) {};
		\node [style=west wire label] (44) at (12.5, -0.75) {reduce-only};
		\node [style=square box] (45) at (9.25, -2) {strong-fert};
		\node [style=west wire label] (46) at (11.25, 2) {rippled};
		\node [style=none] (47) at (-1, -2.5) {};
		\node [style=none] (48) at (2, -2.5) {};
		\node [style=none] (49) at (-1, -1) {};
		\node [style=none] (50) at (2, -1) {};
		\node [style=none] (51) at (15.25, -3.5) {};
		\node [style=none] (52) at (6.75, 1.5) {};
		\node [style=none] (53) at (6.75, -3.5) {};
		\node [style=none] (54) at (15.25, 1.5) {};
	\end{pgfonlayer}
	\begin{pgfonlayer}{edgelayer}
		\draw [style=diredge, in=90, out=-90, looseness=1.25] (17.center) to (7.center);
		\draw [style=diredge, in=90, out=-90, looseness=1.00] (14.center) to (11.center);
		\draw [in=90, out=-90, looseness=1.25] (2.center) to (13.center);
		\draw [style=diredge] (13.center) to (6.center);
		\draw [in=-90, out=-90, looseness=2.00] (5.center) to (10.center);
		\draw (10.center) to (12.center);
		\draw [in=90, out=90, looseness=2.00] (12.center) to (3.center);
		\draw (16.center) to (5.center);
		\draw (9.center) to (17.center);
		\draw (15.center) to (2.center);
		\draw [style=diredge] (3.center) to (1.center);
		\draw [style=diredge] (24.center) to (23.center);
		\draw [style=diredge, in=79, out=-105, looseness=1.50] (25.center) to (26.center);
		\draw [style=diredge] (31.center) to (30.center);
		\draw [style=diredge, in=90, out=-90, looseness=1.00] (38.center) to (33.center);
		\draw [style=diredge, in=90, out=-90, looseness=1.00] (43.center) to (32.center);
		\draw [style=diredge, in=90, out=-90, looseness=1.00] (41.center) to (40.center);
		\draw [style=diredge, in=90, out=-90, looseness=1.00] (35.center) to (39.center);
		\draw [style=dashed] (47.center) to (48.center);
		\draw [style=dashed] (48.center) to (50.center);
		\draw [style=dashed] (50.center) to (49.center);
		\draw [style=dashed] (49.center) to (47.center);
		\draw [style=dashed] (53.center) to (51.center);
		\draw [style=dashed] (51.center) to (54.center);
		\draw [style=dashed] (54.center) to (52.center);
		\draw [style=dashed] (52.center) to (53.center);
		\draw [style=dashed] (50.center) to (52.center);
		\draw [style=dashed] (48.center) to (53.center);
		\draw [in=90, out=-90, looseness=1.00] (34) to (35.center);
	\end{pgfonlayer}
\end{tikzpicture}}
\endpgfgraphicnamed
  \caption{A simple graph hierarchy. The node marked `fertilise' in this PSGraph is itself a PSGraph, consisting of three atomic tactics. \label{fig:unfold-ex}}
\end{figure}

To effectively decide where to send goals, we label edges with \textit{goal-types}, which encode certain properties about a goal which dictate how it should then be handled. We then only allow goals to be output on edges with a matching goal-type. For example, the `induct' tactic might have two output edges: one with type \textit{step}, which matches goals of the form `$?P \implies ?Q$' where $?P$ embeds in $?Q$, and the other with type \textit{base}, which matches everything else. This output types then direct the flow of goals out of the induction tactic in a bigger PSGraph (Figure ~\ref{fig:unfold-ex} (left)).

One evaluation step works by applying a single tactic node on a single goal. Here, the goal is consumed  from the input edge, the tactic in the tactic node is applied to the goal, and the resulting sub-goals (if any) are sent down the output edges where they match. When all the goal nodes are on output edges of the graph, then it has successfully evaluated.
If no output type matches a goal, then evaluation fails. For evaluation this improves robustness of the tactic in two ways: (1) since composition is over the \emph{type of goals}, we avoid the brittleness arising from defining composition in terms of the number of sub-goals or order of sub-goals, and (2) if an unexpected sub-goal arises then evaluation will fail at the actual point of failure as it will not match any of the output types. In general, we allow this evaluation procedure to be non-deterministic by introducing branching whenever a tactic behaves non-deterministically, or a sub-goal produced by a tactic matches more than one output edge. However, with appropriate choice of goal types and evaluation strategy, this branching can be minimised.

Figure~\ref{fig:unfold-ex} (left) highlights an example of a proof strategy employing tactics which rely on specific properties of a goal. For example, \emph{rippling}~\cite{rippling-book} is a heuristic rewriting technique most commonly used on step cases of inductive proofs, ensuring that each `ripple' step moves the goal towards the induction hypothesis (IH). This step is repeated until the IH can be applied to simplify or fully discharge the goal -- a process called `fertilisation'. The advantage of rippling is that it is guaranteed to terminate, whilst allowing rewriting behaviour that would not otherwise terminate (e.g. allowing a rewrite rule to be applied in both directions). Termination is ensured by checking that a certain \emph{embedding} property holds for the goal being rippled, while a measure is reduced from a previous goal. Collectively, these properties are captured by a goal type, in this cased called `\textit{can-ripple}'. When a goal is fully `rippled', then  `fertilisation' is applied.

Proof strategies can easily become very large and complex. In PSGraph, we can
reduce this complexity and size by hiding parts of a graph, which is achieved by
boxing a sub-graph up into a single vertex. We do this by introducing \textit{graph hierarchies}. A simple example of a hierarchy is shown in Figure~\ref{fig:unfold-ex}.

In this paper we extend \cite{LPAR13}, by providing details of the architecture and use of the Tinker tool, which implements the PSGraph formalism. In the next section we exhibit the main aspects of the Tinker UI, namely the means by which users can build and evaluate PSGraphs in the theorem provers Isabelle and ProofPower. In section~\ref{sec:arch}, we provide a more detailed overview of the Tinker architecture, including how to integrate Tinker with a new theorem prover.

\section{Using Tinker}\label{sec:ui}

Currently, Tinker operates in two distinct modes: \textit{evaluation}, where the user employs existing PSGraphs to prove conjectures, and \textit{construction}, where the user builds new PSGraphs. We aim to integrate these two modes in the future, so that strategies can be modified and improved on the fly during evaluation.

\subsection{Evaluating PSGraphs}

PSGraphs are designed to guide, rather than replace, an existing proof system. Thus they should be seen as a generic, theorem prover-independent tool. This is reflected in Tinker, which currently has interfaces implemented for both Isabelle and ProofPower. Users still use the existing interfaces of those provers, with certain extensions provided by Tinker, notably for building PSGraphs and stepping through graph evaluation (see Figure  \ref{fig:arch}). The latter is primarily used in the course of designing or debugging strategies, so Tinker can also evaluate PSGraphs in non-interactive mode, which behaves like normal tactic evaluation.

\subsubsection{Tinkering with Isabelle}

\begin{figure}
\begin{center}
\includegraphics[width=0.7\textwidth]{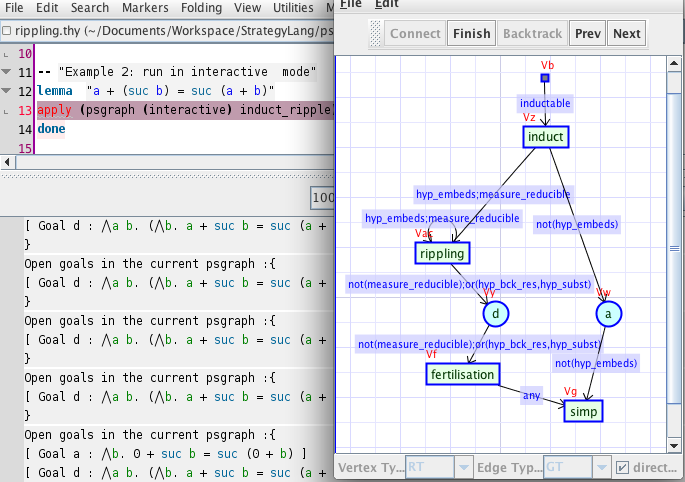}
\end{center}
\vspace{-15pt}
\caption{Tinker GUI with Isabelle}\label{fig:eval:isabelle}
\end{figure}

Tinker is integrated with Isabelle as a new theory on top of the 
`Main' Isabelle/HOL theory\footnote{See \url{https://isabelle.in.tum.de/} for details.}. 
On top of this we have created a new proof method for Isabelle/Isar called
\textsf{psgraph}, which can be applied in one of the following ways:
\begin{tabbing}
\sf\qquad\quad\textbf{apply} (psgraph $\langle$graph-name$\rangle$)\\
\sf\qquad\quad\textbf{apply} (psgraph (interactive) $\langle$graph-name$\rangle$)\\
\sf\qquad\quad\textbf{apply} (psgraph (current))
\end{tabbing}
There are three different modes to work with this tactic. Firstly, if the only argument given is $\langle$graph-name$\rangle$, then it enters the `automatic' mode, which from the user's point of view looks exactly like using any other Isabelle method. The two other modes, `interactive' and `current', utilise the Tinker GUI. This enables users to step through the proof and visualise the flow of goal-nodes. Figure~\ref{fig:eval:isabelle} shows an example of the `interactive' mode using the rippling strategy described above. Here, the related goal information will be printed in the Isabelle output panel. The supported actions in the Tinker GUI are `apply the next box / tactic', `backtrack', `replay the previous step' and `terminate the current evaluation'. If a node represents a nested PSGraph, then a new window is opened showing the nested graph, which the user can evaluate as with the parent graph.

The `current' mode is used when the user builds a new PSGraph in the Tinker GUI, which is described below. Here, the graph that is currently open in the GUI is used by the \textsf{psgraph} method. The available operations in this mode are the same as those in the `interactive' mode.  

PSGraphs are stored in Isabelle's theory context, so to use a PSGraph \textsf{$\langle$graph-name$\rangle$} in `interactive' or `automatic' mode, a graph with the given name first needs to be stored in the theory context.


\subsection{Tinkering with ProofPower}

\begin{figure}
\begin{center}
\centering
\includegraphics[width=0.9\textwidth]{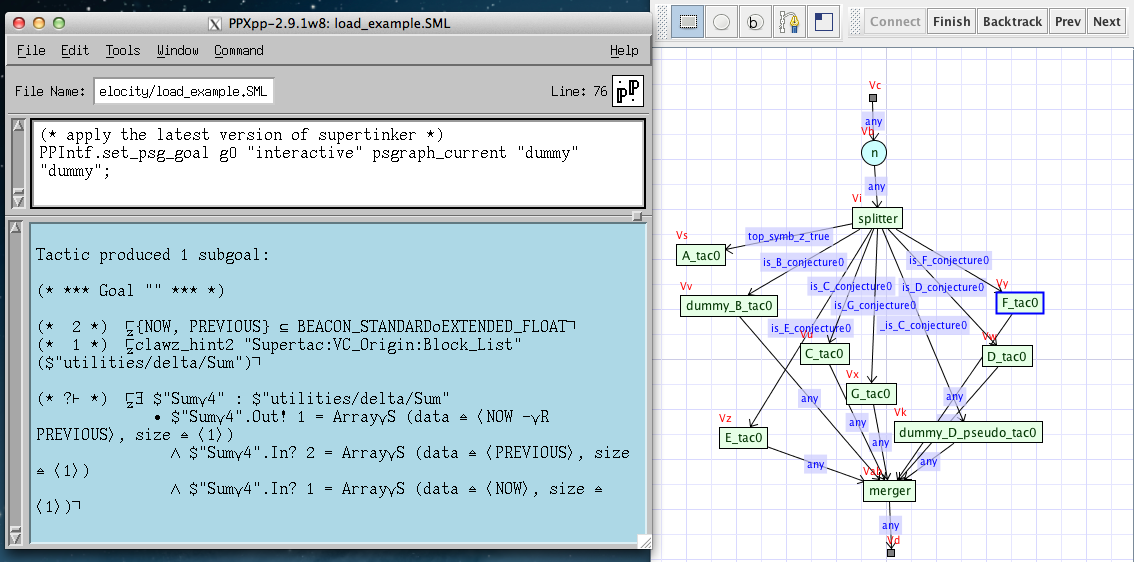}
\end{center}
\vspace{-15pt}
\caption{Tinker GUI with ProofPower}\label{fig:eval:pp}
\end{figure}

The second theorem prover that is supported by Tinker is ProofPower. Here, Tinker has been integrated with ProofPower's subgoal package, used to 
handle goals and soundness of tactics via the kernel. A PSGraph is executed by the function `\textbf{run\_psg\_goal}', which is invoked in a manner similar to the Isabelle method:
\begin{tabbing}
\sf\qquad\quad\textbf{run\_psg\_goal} $\langle$goal$\rangle$ $\langle$graph-name$\rangle$ auto\\
\sf\qquad\quad\textbf{run\_psg\_goal} $\langle$goal$\rangle$ $\langle$graph-name$\rangle$ interactive\\
\sf\qquad\quad\textbf{run\_psg\_goal} $\langle$goal$\rangle$ current
\end{tabbing}
Such calls will initiate a proof of $\langle$goal$\rangle$ with the PSGraph \textsf{$\langle$graph-name$\rangle$}. From this point, Tinker behaves identically to the Isabelle version. Figure~\ref{fig:eval:pp} shows a screenshot of a PSGraph encoding of SuperTac, a powerful and complex ProofPower tactic consisting of thousands lines of ML code and used by D-RisQ\footnote{See \url{www.drisq.com}.} in their ClawZ toolchain \cite{OHalloran13}.

\subsection{Building PSGraphs}

There are effectively two ways to build a PSGraph to be used by the Tinker system. The first is by drawing graphs using the GUI, and the second is by 
programming graphs by combining simpler graphs into more complicated ones, via a handful of \textit{graph combinators}.

\subsubsection{Tinkering by drawing}

\begin{wrapfigure}[11]{r}{0.5\textwidth}
\vspace{-50pt}
\begin{center}
\includegraphics[width=0.5\textwidth]{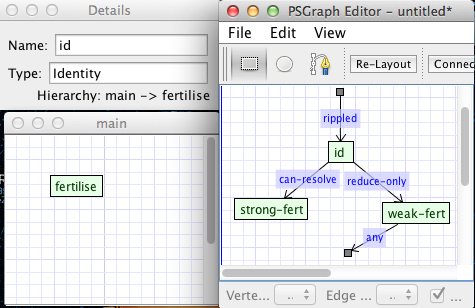}
\end{center}
\vspace{-10pt}
\caption{Tinker Drawing GUI}\label{fig:draw}
\end{wrapfigure}
Drawing a PSGraph in the Tinker GUI is straight forward. Using node and edge tools, the user clicks to place tactic boxes and drags lines to connect boxes with edges. Inputs and outputs are represented using `dummy' vertices (depicted as small grey boxes). Then, double-clicking on nodes or edges allows the user to edit the tactics or goal types, respectively. These tactics and goal types are pre-defined in Tinker, so that they can be looked up by name during evaluation. Also the Tinker GUI allows the user to draw hierarchies as shown in Figure  \ref {fig:draw}. A `Details' window is available to show the name and type of the selected tactics as well as the path of the current hierarchy.

\subsubsection{Tinkering by programming}

In addition to being able to draw new PSGraphs, more complex PSGraphs can be built from simpler ones using a handful of graph combinators.

The atomic tactics available to Tinker correspond to tactics provided by the underlying theorem prover, and a key feature of PSGraph is that it provides a \textbf{type-safe} method to combine the tactics. Graphs are composed by plugging output edges with input edges. These edges are typed and we can only plug ``comparable" types together, where the notion of ``comparable" is goal-type specific. 

Before an atomic tactic can be included in a PSGraph, it must be equipped with a typing on its inputs and outputs. This is achieved by the function \textit{lift}$(\textit{node-nm},\textit{tac},\textit{ins},\textit{outs})$, which produces a graph with a single tactic-node named \textit{node-nm}, with a list of input edges \textit{ins} and output edges \textit{outs}. This tactic-node will then be evaluated by calling the underlying tactic function \textit{tac} in the theorem prover.

\begin{figure}[h]
  \centering
\beginpgfgraphicnamed{combinators}
\begin{tikzpicture}[string graph]
	\begin{pgfonlayer}{nodelayer}
		\node [style=none] (0) at (-6.25, 2.5) {...};
		\node [style=none] (1) at (-5.25, 2.75) {};
		\node [style=cloud vertex, inner sep=2 mm, shape=cloud, aspect=1.5] (2) at (-6.25, 1.25) {$G$};
		\node [style=none] (3) at (-7.25, 2.75) {};
		\node [style=none] (4) at (-6.25, 0) {...};
		\node [style=none] (5) at (-5.25, -2.75) {};
		\node [style=none] (6) at (-6.25, -2.5) {...};
		\node [style=none] (7) at (-7.25, -2.75) {};
		\node [style=cloud vertex, inner sep=2 mm, shape=cloud, aspect=1.5] (8) at (-6.25, -1.25) {$H$};
		\node [style=none] (9) at (-6.25, 3.5) {$G \textrm{ THEN } H$};
		\node [style=east wire label] (10) at (-7.5, -2.75) {$\gamma_1$};
		\node [style=west wire label] (11) at (-5, -2.75) {$\gamma_n$};
		\node [style=west wire label] (12) at (-5, 2.75) {$\alpha_l$};
		\node [style=east wire label] (13) at (-7.5, 2.75) {$\alpha_1$};
		\node [style=west wire label] (14) at (-5, 0) {$\beta_m$};
		\node [style=east wire label] (15) at (-7.5, 0) {$\beta_1$};
		\node [style=cloud vertex, inner sep=2 mm, shape=cloud, aspect=1.5] (16) at (-1.5, 0) {$G$};
		\node [style=none] (17) at (-1.5, 1.25) {...};
		\node [style=none] (18) at (-0.5, -1.5) {};
		\node [style=cloud vertex, inner sep=2 mm, shape=cloud, aspect=1.5] (19) at (2.75, 0) {$G'$};
		\node [style=none] (20) at (-2.5, 1.5) {};
		\node [style=none] (21) at (0.75, 2.25) {$G \textrm{ TENSOR } G'$};
		\node [style=none] (22) at (2.75, 1.25) {...};
		\node [style=none] (23) at (1.75, -1.5) {};
		\node [style=none] (24) at (-0.5, 1.5) {};
		\node [style=none] (25) at (-2.5, -1.5) {};
		\node [style=none] (26) at (3.75, -1.5) {};
		\node [style=none] (27) at (-1.5, -1.25) {...};
		\node [style=none] (28) at (1.75, 1.5) {};
		\node [style=none] (29) at (3.75, 1.5) {};
		\node [style=none] (30) at (2.75, -1.25) {...};
		\node [style=west wire label] (31) at (-0.25, -1.5) {$\beta_m$};
		\node [style=east wire label] (32) at (-2.75, -1.5) {$\beta_1$};
		\node [style=west wire label] (33) at (-0.25, 1.5) {$\alpha_l$};
		\node [style=east wire label] (34) at (-2.75, 1.5) {$\alpha_1$};
		\node [style=west wire label] (35) at (4, 1.5) {$\alpha_l'$};
		\node [style=east wire label] (36) at (1.5, 1.5) {$\alpha_1'$};
		\node [style=west wire label] (37) at (4, -1.5) {$\beta_m'$};
		\node [style=east wire label] (38) at (1.5, -1.5) {$\beta_1'$};
		\node [style=none] (39) at (-4, -3.25) {};
		\node [style=none] (40) at (-8.25, -3.25) {};
		\node [style=none] (41) at (-8.25, 4) {};
		\node [style=none] (42) at (-4, 4) {};
		\node [style=none] (43) at (4.75, -2) {};
		\node [style=none] (44) at (4.75, 2.75) {};
		\node [style=none] (45) at (-3.5, 2.75) {};
		\node [style=none] (46) at (-3.5, -2) {};
		\node [style=cloud vertex, inner sep=2 mm, shape=cloud, aspect=1.5] (47) at (7.25, 0) {$G$};
		\node [style=none] (48) at (8.75, 1.25) {};
		\node [style=none] (49) at (7.25, 1.5) {...};
		\node [style=west wire label] (50) at (10, 0) {$\alpha$};
		\node [style=none] (51) at (6.25, 1.5) {};
		\node [style=none] (52) at (7.25, -1.5) {...};
		\node [style=none] (53) at (8.75, -1.25) {};
		\node [style=none] (54) at (6.25, -1.5) {};
		\node [style=east wire label] (55) at (6, 1.5) {$\gamma_1$};
		\node [style=east wire label] (56) at (6, -1.5) {$\beta_1$};
		\node [style=none] (57) at (8, 2.25) {$\textrm{REPEAT}_\alpha(G)$};
		\node [style=none] (58) at (10.5, 2.75) {};
		\node [style=none] (59) at (5.25, 2.75) {};
		\node [style=none] (60) at (5.25, -2) {};
		\node [style=none] (61) at (10.5, -2) {};
	\end{pgfonlayer}
	\begin{pgfonlayer}{edgelayer}
		\draw [style=diredge, in=120, out=-90, looseness=1.00] (3.center) to (2);
		\draw [style=diredge, in=60, out=-90, looseness=1.00] (1.center) to (2);
		\draw [style=diredge, in=90, out=-124, looseness=1.00] (8) to (7.center);
		\draw [style=diredge, in=90, out=-56, looseness=1.00] (8) to (5.center);
		\draw [style=diredge, bend right=45, looseness=1.00] (2) to (8);
		\draw [style=diredge, bend left=45, looseness=1.00] (2) to (8);
		\draw [style=diredge, in=120, out=-90, looseness=1.00] (20.center) to (16);
		\draw [style=diredge, in=60, out=-90, looseness=1.00] (24.center) to (16);
		\draw [style=diredge, in=90, out=-60, looseness=1.00] (16) to (18.center);
		\draw [style=diredge, in=90, out=-120, looseness=1.00] (16) to (25.center);
		\draw [style=diredge, in=120, out=-90, looseness=1.00] (28.center) to (19);
		\draw [style=diredge, in=60, out=-90, looseness=1.00] (29.center) to (19);
		\draw [style=diredge, in=90, out=-60, looseness=1.00] (19) to (26.center);
		\draw [style=diredge, in=90, out=-120, looseness=1.00] (19) to (23.center);
		\draw [style=gray edge] (41.center) to (42.center);
		\draw [style=gray edge] (42.center) to (39.center);
		\draw [style=gray edge] (39.center) to (40.center);
		\draw [style=gray edge] (40.center) to (41.center);
		\draw [style=gray edge] (45.center) to (44.center);
		\draw [style=gray edge] (44.center) to (43.center);
		\draw [style=gray edge] (43.center) to (46.center);
		\draw [style=gray edge] (46.center) to (45.center);
		\draw [style=diredge, in=120, out=-90, looseness=1.00] (51.center) to (47);
		\draw [style=diredge, in=90, out=-120, looseness=1.00] (47) to (54.center);
		\draw [in=0, out=0, looseness=1.25] (53.center) to (48.center);
		\draw [style=diredge, in=45, out=180, looseness=0.75] (48.center) to (47);
		\draw [in=180, out=-45, looseness=0.75] (47) to (53.center);
		\draw [style=gray edge] (59.center) to (58.center);
		\draw [style=gray edge] (58.center) to (61.center);
		\draw [style=gray edge] (61.center) to (60.center);
		\draw [style=gray edge] (60.center) to (59.center);
	\end{pgfonlayer}
\end{tikzpicture}}
\endpgfgraphicnamed
    \vspace{-7pt}
  \caption{\label{fig:then-and-tensor} THEN, TENSOR, and $\textrm{REPEAT}_\alpha$ combinators}
\end{figure}
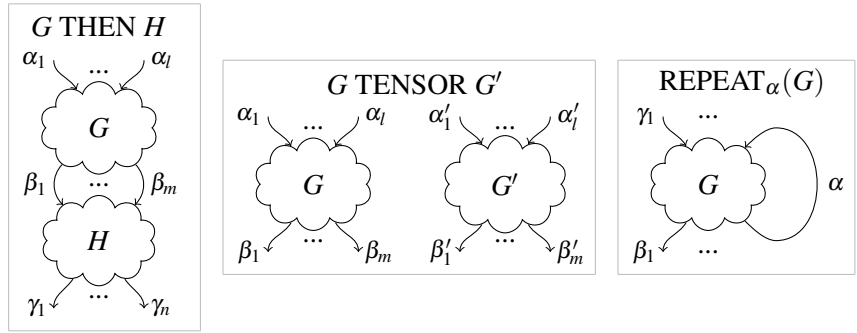

One of the most common ways to combine tactics is via a \emph{THEN} tactical, where $t_1~\textit{THEN}~t_2$ first applies tactic $t_1$ then tactic $t_2$. PSGraphs have a corresponding combinator, also called \textit{THEN}, where $G~\textit{THEN}~H$ plugs each output edge of $G$ into an input edge of $H$ 
of the same type. This is illustrated on the left-most diagram of Figure  \ref{fig:then-and-tensor}. Note that there could be multiple ways of plugging PSGraphs together, so Tinker provides two versions of the combinator: one version that returns a list of all possible ways $G$ could be plugged into $H$ and one version that simply picks one. Further, note that a \emph{maximal plugging} is always done, meaning if two edges can be connected they will be. Any remaining, un-plugged inputs/outputs will become inputs/outputs of the new graph.

Another combinator is \textit{TENSOR}, which puts $G$ and $G'$ side by side in a single graph, without doing any plugging. This is depicted in the middle diagram of Figure~\ref{fig:then-and-tensor}. The right-most diagram shows how repetition can be naturally represented as a feedback edge. This is can be achieved by invoking the $\textit{REPEAT}_\alpha$ combinator, which plugs an output of goal type $\alpha$ to an input of the same type. If we think of $\alpha$ as a loop-condition, this provides a graphical analogue to the common \emph{REPEAT\_WHILE} or \emph{REPEAT\_UNTIL} tacticals. While it is possible to encode the more common \emph{REPEAT} tactical\footnote{To be precise, repeating $t$ until it fails can be encoded using the existing combinators as:
\vspace{-4pt}
$$
\textit{REPEAT}_\alpha\Big(
\big(\textit{lift}(t,t,[\alpha,\alpha][\alpha]) ~\textit{TENSOR}~
\textit{lift}(id,id,[],[\beta])\big)~\textit{ORELSE}~
\big(\textit{lift}(id,id,[\alpha,\alpha][\beta]) ~\textit{TENSOR}~
\textit{lift}(id,id,[],[\alpha])\big)\Big)
$$
\vspace{-12pt}

\noindent where $id$ is the identity tactic, and $\alpha$ and $\beta$ are goal types that succeeds for any goal.} we do not see this as a natural way of working with PSGraphs, as the goal is make a proof strategy developer think about why certain choices have been made, and represent this intuition in the goal types.

\begin{figure}[h]
  \centering
  $\textit{NEST}(\textrm{g-tac}, G) \ =\  %
\beginpgfgraphicnamed{nest_def}
\begin{tikzpicture}[string graph]
	\begin{pgfonlayer}{nodelayer}
		\node [style=none] (0) at (-4.5, -1) {};
		\node [style=none] (1) at (-1.5, -1) {};
		\node [style=none] (2) at (-4.5, 1) {};
		\node [style=none] (3) at (-1.5, 1) {};
		\node [style=none] (4) at (7.75, -2.5) {};
		\node [style=none] (5) at (1.5, 2.5) {};
		\node [style=none] (6) at (1.5, -2.5) {};
		\node [style=none] (7) at (7.75, 2.5) {};
		\node [style=none] (8) at (4.5, 2) {...};
		\node [style=west wire label] (9) at (6.75, -3) {$\beta_m$};
		\node [style=none] (10) at (6.5, -3) {};
		\node [style=none] (11) at (2.5, 3) {};
		\node [style=west wire label] (12) at (6.75, 3) {$\alpha_l$};
		\node [style=east wire label] (13) at (2.25, -3) {$\beta_1$};
		\node [style=east wire label] (14) at (2.25, 3) {$\alpha_1$};
		\node [style=none] (15) at (2.5, -3) {};
		\node [style=cloud vertex, inner sep=4 mm, aspect=1.5] (16) at (4.5, 0) {$G$};
		\node [style=none] (17) at (6.5, 3) {};
		\node [style=none] (18) at (4.5, -2) {...};
		\node [style=west wire label] (19) at (-1.75, -1.5) {$\beta_m$};
		\node [style=none] (20) at (-2, -1.5) {};
		\node [style=square box, inner sep=2 mm, aspect=1.5] (21) at (-3, 0) {g-tac};
		\node [style=none] (22) at (-2, 1.5) {};
		\node [style=east wire label] (23) at (-4.25, -1.5) {$\beta_1$};
		\node [style=none] (24) at (-4, -1.5) {};
		\node [style=none] (25) at (-3, -1.25) {...};
		\node [style=none] (26) at (-4, 1.5) {};
		\node [style=west wire label] (27) at (-1.75, 1.5) {$\alpha_l$};
		\node [style=east wire label] (28) at (-4.25, 1.5) {$\alpha_1$};
		\node [style=none] (29) at (-3, 1.25) {...};
	\end{pgfonlayer}
	\begin{pgfonlayer}{edgelayer}
		\draw [style=dashed] (0.center) to (1.center);
		\draw [style=dashed] (1.center) to (3.center);
		\draw [style=dashed] (3.center) to (2.center);
		\draw [style=dashed] (2.center) to (0.center);
		\draw [style=dashed] (6.center) to (4.center);
		\draw [style=dashed] (4.center) to (7.center);
		\draw [style=dashed] (7.center) to (5.center);
		\draw [style=dashed] (5.center) to (6.center);
		\draw [style=dashed] (3.center) to (5.center);
		\draw [style=dashed] (1.center) to (6.center);
		\draw [style=diredge, in=120, out=-90, looseness=1.00] (11.center) to (16);
		\draw [style=diredge, in=60, out=-90, looseness=1.00] (17.center) to (16);
		\draw [style=diredge, in=90, out=-60, looseness=1.00] (16) to (10.center);
		\draw [style=diredge, in=90, out=-120, looseness=1.00] (16) to (15.center);
		\draw [style=diredge, in=120, out=-90, looseness=1.00] (26.center) to (21);
		\draw [style=diredge, in=60, out=-90, looseness=1.00] (22.center) to (21);
		\draw [style=diredge, in=90, out=-60, looseness=1.00] (21) to (20.center);
		\draw [style=diredge, in=90, out=-120, looseness=1.00] (21) to (24.center);
	\end{pgfonlayer}
\end{tikzpicture}}
\endpgfgraphicnamed$
  \vspace{-7pt}
    \caption{\label{fig:nest} NEST combinator, which collapses a graph $G$ to a single node `g-tac'}
\end{figure}
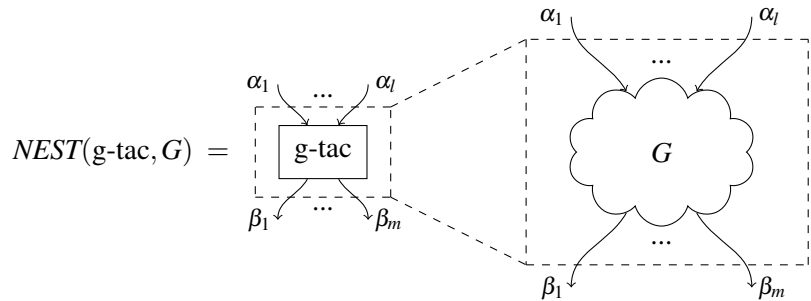

Hierarchies are supported in PSGraph, and the \textit{NEST} combinator introduces a hierarchy by ``collapsing" the graph into a single graph tactic with the original graph nested. This is illustrated graphically in Figure  \ref{fig:nest}. Two generalisations of \textit{NEST} are the \emph{OR} and \emph{ORELSE} combinators, which rather than nesting a single graph, will nest a pair of graphs. When evaluating $G~\emph{OR}~H$, Tinker will branch, trying to evaluate both $G$ and $H$. On the other hand, $G~\emph{ORELSE}~H$ will only evaluate $H$ if the evaluation of $G$ fails. In the current version of Tinker, $G$ and $H$ must have the exact same number and types of input and output edges, however a direct generalisation is to produce a nested graph node whose inputs are the intersection of the inputs of $G$ and $H$, and whose outputs are the union of the outputs of $G$ and $H$.

\section{Tinker tool architecture}\label{sec:arch}

\begin{figure}[h]
\vspace{-15pt}
\begin{center}
\includegraphics[width=0.64\textwidth]{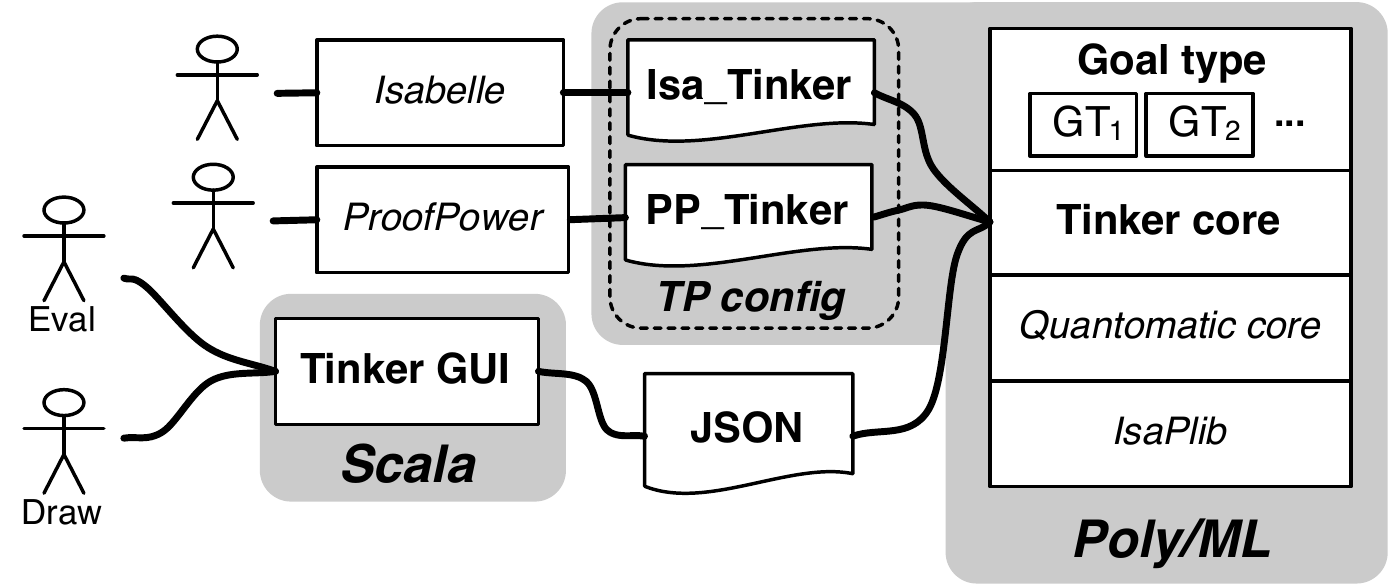}
$\qquad$
\includegraphics[width=0.3\textwidth]{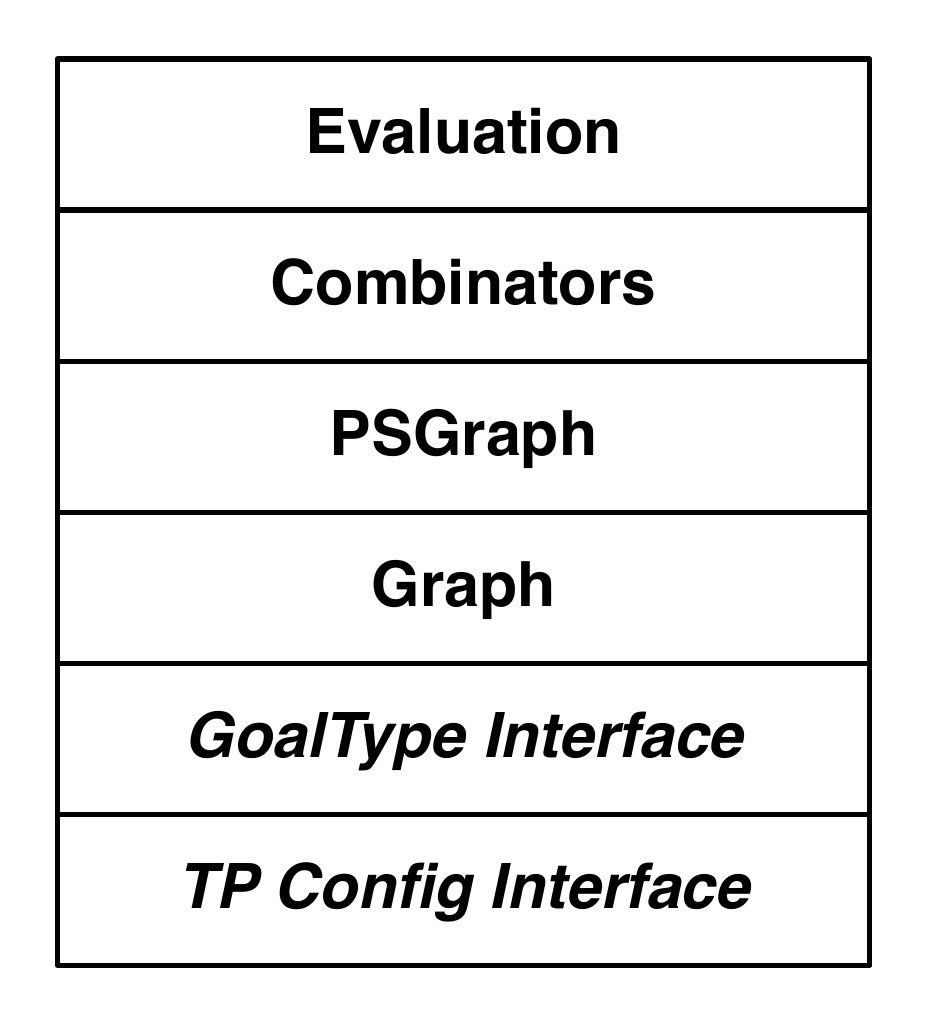}
\end{center}
\vspace{-15pt}
\caption{The Tinker architecture (left) and the Tinker core components (right)}\label{fig:arch}
\end{figure}

\noindent Figure \ref{fig:arch} (left) shows the architecture of the tool. The tool consists of two overall parts, each are in separate grey boxes.
The \textbf{core} is where all the ``magic'' happens, and is implemented in Poly/ML. The \textbf{GUI} is implemented in Scala and provides
a graphical way of creating, modifying, evaluating and debugging PSGraphs. The Tinker tool is independent of the underlying theorem prover and goal types. In both cases, this extra data is provided via ML structures passed to the Tinker core functors.

Figure \ref{fig:arch} (right) shows the main components of the Tinker core, and their dependencies, and each component is detailed below.

\subsection{IsaPlib \& the Quantomatic core}\label{sec:quanto}

The Tinker tool started out as a new language to write proof plans in the IsaPlanner proof planner \cite{paper:Dixon:03} for Isabelle. 
However, Tinker has evolved considerably since then, and the only parts used are those  
 retained in an ML library called IsaPlib, which can either co-exist with Isabelle/ML or provide an Isabelle/ML-like environment and adds some extra functionality.


PSGraphs themselves are open-graphs, and evaluation is done by open-graph rewriting. A formalisation and rewrite theory for open-graphs was provided in~\cite{paper:Dixon:10}. The Quantomatic core~\cite{Quantomatic} provides a general purpose open-graph rewriting library on top of IsaPlib, which is employed by Tinker for PSGraph evaluation.

\subsection{The Tinker GUI \& JSON communication protocol}

\begin{wrapfigure}[8]{r}{0.55\textwidth}
\vspace{-15pt}
\begin{center}
\includegraphics[width=0.5\textwidth]{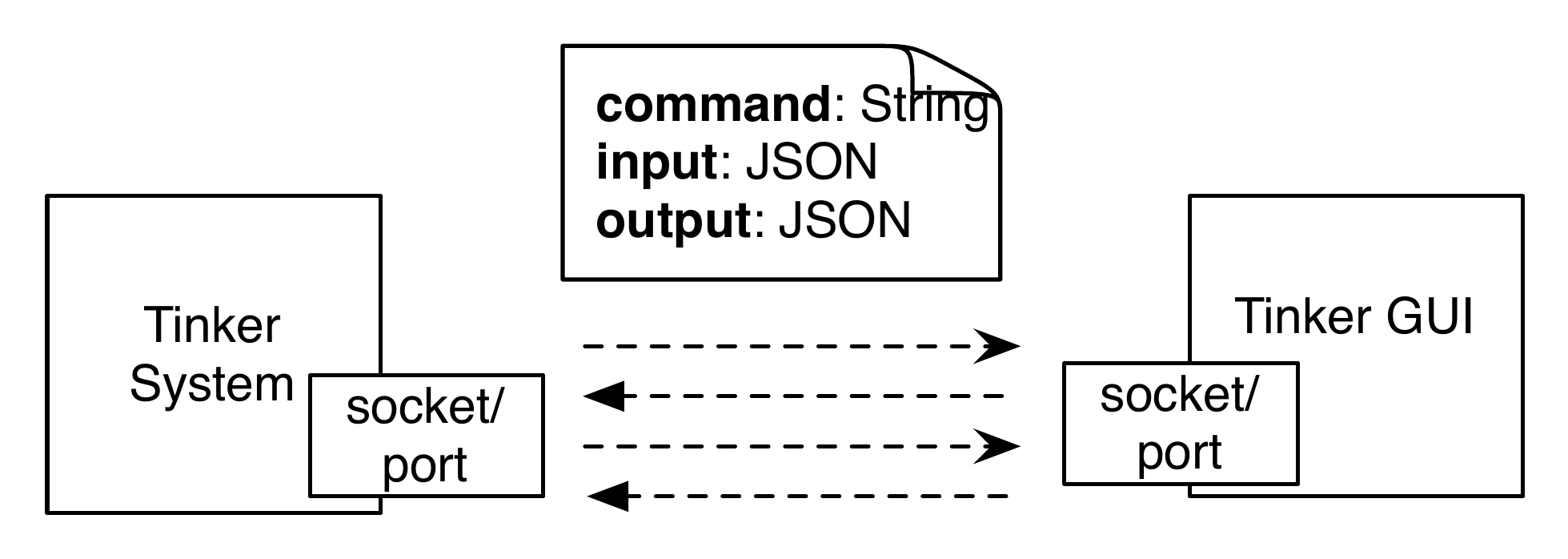}
\end{center}
\vspace{-15pt}
\caption{Communication with JSON protocol}\label{fig:json protocol}
\end{wrapfigure}

\noindent A JSON protocol is used to communicate between the Scala GUI and Poly/ML core. 
As seen in Figure \ref{fig:json protocol}, the GUI communicates with the core over a socket in a client/server mode.
Along with a unique message ID, a JSON message has three relevant fields: \textit{command}, \textit{input} and \textit{output}. 
The command is a string indicating a command defined by the core. 
Both input and output are in the JSON format, thus 
data that needs to be communicated between the core and the GUI must always be serialisable as JSON.

The protocol is implemented on top of the interface provided by Quantomatic, with Tinker specific features to work with graphs (such as evaluation and hierarchies).

\subsection{Goal types}

There are two perspectives one can take on goal types. The first, which was presented in~\cite{LPAR13}, is to view a goal type simply as a \textit{predicate} on a goal. In this sense, typing provides a mechanism for filtering goals as they are output from tactics to send them to the right place. A goal type also encodes a strategy-writer's expectation for what goals should look like at various points throughout the proof. This can help to make sense of large and complex proof strategies, and more importantly, \emph{why} and \emph{where} an evaluation has failed. This information can then be used to patch the proof strategy, either manually, or automatically via a \emph{proof critic} mechanism (c.f. \cite{rippling-book}). 

The second perspective on goal types, which subsumes the first, is that they should be used a general mechanism for \emph{guiding} proofs and tactics. Here, instead of just seeing a goal as a predicate it is seen as
a (partial) \emph{function} on goal nodes. It is undefined in cases where the type does not match the goal, but when it does match, this function can be used to augment the goal-node with additional information, which can for example highlight relevant lemmas/hypothesis for a tactic. This data can in turn be used later by other tactics. Such goal types can even be initialised automatically via machine-learning techniques. A first approach to this is discussed in~\cite{grov13a}.  Tinker supports this more general kind of goal type, but for the sake of simplicity, we have chosen to focus mainly on `predicate-style' types for this paper.

\subsubsection{Creating a new goal type -- the GoalType interface}

To create a new goal type the GoalType signature has to be implemented. Firstly, types
\begin{verbatim}
  type T
  type gnode
\end{verbatim}
to represent a goal type and a goal node on the graph must be implemented, as well as a matching function:
\begin{verbatim}
  val match :  gnode -> T -> pnode -> gnode option
\end{verbatim}
where \texttt{match} is the functional view of a goal type, and a predicate will just return the same \texttt{gnode} for success and \texttt{NONE} if it fails.
Note that \texttt{gnode} is the goal node on the graph, while \texttt{pnode} is an element of the underlying proof representation. The key differences between them is that a \texttt{gnode} is on the graph and must be serialisable, while this is not required for a \texttt{pnode}. Note that the  \texttt{gnode}, is the ``previous node" (the one that generated the \texttt{pnode}) and is used to encode features such as `measure-reduction' for rippling. We will return to \texttt{pnode}s below.

For the `basic' goal type \cite{LPAR13}, \texttt{T} is a string. The string is assumed to have a set of predicates separated by `;', and for each of these names a matching function has to be created, which is called by \texttt{match}. One example is  `top\_symbol'. When the substring `top\_symbol' is found then a matching function is called that checks if the top level symbol of the goal matches with a given term. This implementation of goal-types has limited support for logic in the form of `not' and `or' connectives. To work with the GUI, a translation between \texttt{T} and \texttt{gnome} and a JSON representation must also be provided.

\subsection{Graph \& PSGraph representation}

As we mentioned in Section~\ref{sec:quanto}, Tinker uses the Quantomatic library for graph rewriting. This is done by instantiating a \textit{GRAPHICAL\_THEORY} structure called \textit{PSGraph\_Theory}. To instantiate a graphical theory, types for graph nodes and edges need to be provided, along with relevant functions for data matching and I/O. Quantomatic then constructs structures for graphs, graph rewrite rules, matching, rewriting, etc.

For the edges, the data is the \emph{goaltype} already discussed. The nodes are either tactic-nodes, whose data is represented by a \textit{reasoning technique} record, or goal-nodes. The reasoning technique record looks like this:
{\it \begin{tabbing}
$\qquad\quad$\textit{RTechn} := \{ \= name :: name , appf :: app\_data \} \\
\end{tabbing}}
\vspace{-15pt}
\noindent A tactic-node is evaluated according to the contents of the \emph{app\_data} field:
{\it \begin{tabbing}
$\qquad\quad$\textbf{type} app\_data := Tactic name $|$ Nested (name,\{Or,OrElse\}) \\
\end{tabbing}}
\vspace{-15pt}
\noindent 
This field contains the name of an atomic tactic (i.e. a tactic provided by the theorem prover), or the name of a graph tactic, which corresponds to a list of nested PSGraphs. In the latter case, a flag is used to indicate whether multiple nested graphs should be evaluated OR-style or ORELSE-style.
The data on goal-nodes is given by the \texttt{gnode} type, described before.

We define a \emph{PSGraph} as follows:
\newcommand{\fmap}{\mbox{$~\stackrel{m}{\rightarrow}~$}}
{\it
\begin{tabbing}
\quad\textit{PSGraph} := \{ \= graph :: Graph, 
 atomics :: \textit{name} \fmap{} \textit{tactic},
 graph\_tactics :: name \fmap{} [ Graph ] \} \\
\end{tabbing}}
\vspace{-15pt}
\noindent 
The \emph{graph} field contains the graph of the proof strategy. The \emph{atomics} field is a map from names of atomic tactics to actual tactic functions provided by the prover. The \emph{graph\_tactics} field contains any nested graph tactics.

\subsection{Combinators revisited}

With the underlying PSGraph representation, we can provide more details about how the combinators are implemented. Since PSGraphs all contain their own lookup tables for atomic/graph tactics, it is important to avoid name clashes when we combine multiple graphs into one. To avoid such clashes we define a function type
$$
\textbf{type}~\textit{psgraph\_fun} = \textit{PSGraph} \rightarrow \textit{PSGraph}
$$
which is then used to combine PSGraphs. The idea is that given a PSGraph we generate fresh names when required, and this type enables us to linearise this generation by enforcing an order. The combinators then have types given by:
$$
\begin{array}{rcl}
\textit{lift} & :: & \textit{name} \times \textit{tactic} \times [\textit{goaltype}] \times [\textit{goaltype}] \rightarrow \textit{psgraph\_fun} \\
\textit{THEN} & :: & \textit{psgraph\_fun} \times \textit{psgraph\_fun} \rightarrow \textit{psgraph\_fun} \\
\textit{TENSOR} & :: & \textit{psgraph\_fun} \times \textit{psgraph\_fun} \rightarrow \textit{psgraph\_fun} \\
\textit{REPEAT} & :: & \textit{psgraph\_fun} \times \textit{goaltype} \rightarrow \textit{psgraph\_fun} \\
\textit{NEST} & :: & \textit{name} \times \textit{psgraph\_fun} \to  \textit{psgraph\_fun} \\
\textit{OR} & :: &  \textit{name} \times \textit{psgraph\_fun} \times \textit{psgraph\_fun} \rightarrow \textit{psgraph\_fun} \\
\textit{ORELSE} & :: &  \textit{name} \times \textit{psgraph\_fun} \times \textit{psgraph\_fun} \rightarrow \textit{psgraph\_fun}
\end{array}
$$

A \textit{psgraph\_fun} can be turned into a PSGraph by applying it to the empty graph:
$$
\textit{empty\_psgraph} :: \textit{PSGraph}
$$

\subsection{The TP interface... OR how to connect a new theorem prover to Tinker}

To connect a theorem prover, a structure implementing the \emph{TP Config} signature must be provided. These structures for Isabelle and ProofPower are depicted in Figure ~\ref{fig:arch} as \emph{Isa\_Tinker} and \emph{PP\_Tinker}. This structure contains all theorem prover related data types and functions that are required. Firstly, data types for the underlying terms, theorems, proof context and tactics must be provided\footnote{The list is not complete but gives an indication of the interface between Tinker and the underlying theorem prover.}:
\begin{verbatim}
  type term 
  type thm
  type context
  type tactic 
\end{verbatim}
A handful of common functions over these types must also be provided, such as matching of terms:
\begin{verbatim}
  val match : context -> (term * term) -> bool
\end{verbatim}
Tinker assumes that a (partial) proof will be of type \texttt{plan} and it contains a set of open goals of type \texttt{pnode}. A tactic
is then seen as a function from a (partial) proof (\texttt{pplan}) and a goal (\texttt{pnode}) to a list of new goals and 
an updated proof. As there may be multiple branches when applying a tactic, a sequence of each branch is returned. This is captured by 
the \texttt{appf} type:
\begin{verbatim}
  type pplan
  type pnode
  type appf = pnode * pplan -> (pnode list * pplan) Seq.seq
\end{verbatim}
The user also needs to provide a means of producing \texttt{appf} functions from tactics, which are allowed to take a list of theorems as input:
\begin{verbatim}
  val apply_tactic : thm list -> tactic -> appf
\end{verbatim}

This interface was designed to provide the necessary functionality to interface with both Isabelle and ProofPower. We expect these to evolve when connecting more theorem provers in the future. Also note that so far both connected theorem provers are implemented in Poly/ML, making integration easier. For other platforms, integration may need to be done using e.g. socket communication.

\vspace{-5pt}
\subsection{Evaluating PSGraphs}
\vspace{-5pt}

Evaluation makes heavy use of both the prover and goal type functionality provided. A special record \textit{Eval} is used to keep track of the evaluation state of a PSGraph $G$ as it applied to a \texttt{pnode} $p$ of a \texttt{plan} \textit{prf}. \textit{Eval} contains $G$ itself, together with a list of branches of the evaluation. Each branch contains a stack of active graphs (to support hierarchical evaluation of PSGraphs) as well as \textit{prf}.

In general, the graph $G$ may have more than one input edge, so we create a branch for each input edge $e$ by placing a goal-node containing $p$ on $e$ -- as long as this goal-node matches the goal type of $e$.

Evaluation follows by picking a branch $b$ and searching the PSGraph at the top of $b$'s stack for a goal node $g$ situated on an input edge of a tactic-node $t$. Firstly, $g$ is removed from the graph. Then if $t$ is an atomic tactic,  it is applied to $g$, using the \texttt{apply\_tactic} function. 
As a tactic may introduce branching, a sequence of updated \texttt{pplan}s coupled with a list $ps$ of newly generated subgoals (\texttt{pnode}s), is returned.  For each element in this sequence, the new subgoals are added to the output edge of $t$ in a type-respecting way, without duplicating or loosing any of the subgoals. There may be multiple ways of achieving this, and each such way becomes a branch. It may also not be possible, which 
means that this element is discarded. To add a \texttt{pnode} to an edge, it must first be converted to a (serialisable) \texttt{gnode}. Such \texttt{gnode} is the return value of the \texttt{match} function of the goal type when it succeeds.

If $t$ is a graph tactic, then $g$ is moved to the corresponding input edge of the graph nested inside of $t$, and this graph is pushed on to the stack of active PSGraphs. A graph evaluation has successfully terminated when all goal nodes are on the output edges of the graph. If this is the only graph on the stack then the evaluation is complete. Otherwise, the top graph is popped from the stack and its output goals are added back to the parent graph on the appropriate edges.

Graph tactics nesting multiple graphs via \textit{OR} or \textit{ORELSE} are evaluated similarly, except branches are created for each subgraph, which are all evaluated (for \textit{OR}) or evaluated until one branch completes successfully (for \textit{ORELSE}).

\vspace{-15pt}
\section{Conclusion, related \& future work}
\vspace{-10pt}

In this paper we have extended \cite{LPAR13}, which introduced the PSGraph formalisation, with details of how PSGraphs have been implemented in the Tinker tool -- with support for the Isabelle and ProofPower theorem provers. 

There are other tools supporting graphical representation of proofs. L$\Omega$UI \cite{SiekmannHBCFHKKMMPS99} and XIsabelle \cite{Ozols97} enable a graphical view of proof trees. They deviate from Tinker by viewing the proof and not the underlying proof strategies, which requires e.g. loops. Moreover, they cannot be used for inspection of the goal flow during evaluation, and crucially, the do not have the necessary goal abstraction via goal types. Moreover, tools such as ProveEasy \cite{Burstall2000} and
Jape \cite{bornat1997jape} have been developed to help students to learn how to do formal proofs, and deviates from Tinker in the same way as XIsabelle and L$\Omega$UI. Finally, as with XBarnacle\cite{Lowe97xbarnacle},  Tinker supports hierarchies, thus one can view a proof (strategy) at different levels of abstraction. Incorporating other features from XBarnacle is future work.

This is the first version of the tool, and we have already started the work on the next version, and we will now summarise features we would like to include in the next version. Firstly, goal types are a topic of active research, both from a theoretical perspective and from the point of view of implementation. The tool has therefore been developed to be flexible in the sense we can easily plug in new goal types. Experiments so far have shown that the predicate-style types described in~\cite{LPAR13} are easy to work with, but do not allow certain desirable features. On the other hand, the more expressive types from~\cite{grov13a} are very powerful, but complex to work with.

Therefore, we are now focusing on defining a new goal type combining the simplicity of~\cite{LPAR13} with the power of~\cite{grov13a}, extended with variables and pattern matching as in e.g. {\cal L}tac \cite{Delahaye02}.

In the next version of Tinker, we would like to integrate the evaluation and development views of the GUI to enable changing graphs during evaluation.
The GUI should also show the proof state, and offer an improved layout mechanism whereby only goal-nodes change position during the course of evaluation. Moreover, more user control during evaluation is desirable, in particular allowing users to easily select which goal should be executed next, and combine `interactive' and `automatic' modes, where the user can choose to `step over' or `step into' hierarchies. Continuing with this `debugger' paradigm, we would also like to add breakpoints in order to start `interactive' mode at a particular step during an otherwise `automatic' evaluation. This will be very useful for debugging large tactics such as SuperTac~\cite{OHalloran13}.

For the building aspect, we would like to make it easier to fold/unfold graphs when building new graphs, combined with a distributed way of defining search and evaluation strategies, where each nested graph may have its own strategy -- configurable in the GUI. We would also like to add `Library' functionality, as in e.g. Simulink\footnote{See \url{www.mathworks.co.uk/products/simulink/}.}, where a user can drag-and-drop existing, pre-made strategies into the current graph or create new Library entries from re-usable graph components.

Finally, we plan to integrate Tinker with more theorem provers. HOL4 should provide fewest obstacles as it is close to ProofPower and written in PolyML. Other provers, such as HOL light and Rodin, will require socket communication between Tinker and the prover core.

\vspace{-15pt}

\begin{footnotesize}
\bibliographystyle{eptcs}
\bibliography{stratlang}
\end{footnotesize}
\end{document}